\pdfoutput=1

\documentclass[12pt,preprint]{aastex631}
\usepackage{amsmath}
\usepackage{graphicx}
\usepackage{mathrsfs}
\usepackage{color}
\usepackage{url}
\usepackage{CJKutf8}
\usepackage{ulem}
\usepackage{wasysym}
\usepackage{multirow}
\usepackage{booktabs}

\received{2024}
\revised{\today}
\accepted{2024}

\shorttitle{Saturn Trojan 2019 UO$_{14}$}
\shortauthors{Hui et al. 2024}

\begin{document}

\title{
2019 UO$_{14}$: A Transient Trojan of Saturn
}

\correspondingauthor{Man-To Hui}
\email{mthui@must.edu.mo, manto@hawaii.edu}

\author{\begin{CJK}{UTF8}{bsmi}Man-To Hui (許文韜)\end{CJK}}
\affiliation{State Key Laboratory of Lunar and Planetary Sciences, 
Macau University of Science and Technology, 
Avenida Wai Long, Taipa, Macau}

\author{Paul A. Wiegert}
\affiliation{Department of Physics and Astronomy,
The University of Western Ontario,
1151 Richmond Street,
London, ON N6A 3K7, Canada}
\affiliation{Institute for Earth and Space Exploration,
The University of Western Ontario,
1151 Richmond Street,
London, ON N6A 3K7, Canada}

\author{Robert Weryk}
\affiliation{Department of Physics and Astronomy,
The University of Western Ontario,
1151 Richmond Street,
London, ON N6A 3K7, Canada}

\author{Marco Micheli}
\affiliation{ESA NEO Coordination Centre, Planetary Defence Office,
Largo Galileo Galilei 1, I-00044 Frascati (RM), Italy}

\author{David J. Tholen}
\affiliation{Institute for Astronomy,
University of Hawai`i,
2680 Woodlawn Drive, Honolulu, HI 96822, USA}

\author{Sam Deen}
\affiliation{Simi Valley, CA, USA}

\author{Andrew J. Walker}
\affiliation{Tuggeranong, ACT, Australia}

\author{Richard Wainscoat}
\affiliation{Institute for Astronomy, 
University of Hawai`i, 
2680 Woodlawn Drive, Honolulu, HI 96822, USA}


\begin{abstract}

Saturn has long been the only giant planet in our solar system without any known Trojan members. In this paper, with serendipitous archival observations and refined orbit determination, we report that 2019 UO$_{14}$ is a Trojan of the gas giant. However, the object is only a transient Trojan currently librating around the leading Lagrange point $L_4$ of the Sun-Saturn system in a period of $\sim\!0.7$ kyr. Our N-body numerical simulation shows that 2019 UO$_{14}$ was likely captured as a Centaur and became trapped around $L_4$ $\sim\!2$ kyr ago from a horseshoe coorbital. The current Trojan state will be maintained for another millennium or thereabouts before transitioning back to a horseshoe state. Additionally, we characterize the physical properties of 2019 UO$_{14}$. Assuming a linear phase slope of $0.06 \pm 0.01$ mag/deg, the mean $r$-band absolute magnitude of the object was determined to be $H_r = 13.11 \pm 0.07$, with its color measured to be consistent with those of Jupiter and Neptune Trojans and not statistically different from Centaurs. Although the short-lived Saturn Trojan exhibited no compelling evidence of activity in the observations, we favour the possibility that it could be an active Trojan. If confirmed, 2019 UO$_{14}$ would be marked as the first active Trojan in our solar system. We conservatively determine the optical depth of dust within our photometric aperture to be $\la\!10^{-7}$, corresponding to a dust mass-loss rate to be $\la\!1$ kg s$^{-1}$, provided that the physical properties of dust grains resemble Centaur 29P/Schwassmann-Wachmann 1.

\end{abstract}

\keywords{
Asteroids; Trojan asteroids
}

\section{Introduction}
\label{sec_intro}

Trojans are a class of small bodies that are trapped and librate around the leading ($L_4$) or trailing ($L_5$) Lagrange points of some Sun-planet system. They were either formed in situ as primordial bodies since the planetary formation or were captured from elsewhere. To date, over 13,000 Jupiter Trojans, one Uranus Trojan, and 31 Neptune Trojans have been discovered. However, Saturn, in spite of being the second most massive planet in the solar system and bearing many similarities to Jupiter, is the only giant planet without any discovered Trojans.\footnote{See the list of known Trojans at \url{https://minorplanetcenter.net//iau/lists/Trojans.html}.} The peculiar absence of Saturn Trojans is attributed to dynamical removal during the planetary migration phase \citep{1998AJ....116.2590G}, destruction by mutual collisions \citep{1996Icar..119..192M,1997Icar..125...39M}, and/or the stable regions around $L_4$ and $L_5$ of the Sun-Saturn system being much smaller than those the Sun-Jupiter system \citep[e.g.,][]{1993AJ....105.1987H,2014MNRAS.437.1420H} due to perturbations from the near 5:2 mean-motion resonance between Jupiter and Saturn \citep{1989AJ.....97..900I,1996Icar..121...88D,2002Icar..160..271N} and/or the presence of secular resonances \citep{2000Icar..146..232M,2002ApJ...579..905M}. Nevertheless, it has been suggested that stable regions around $L_4$ and $L_5$ of the Sun-Saturn system may exist, which opens up the possibility for the discovery of a small population of Saturn Trojans \citep{1996Icar..121...88D,2001MNRAS.322L..17M,2002ApJ...579..905M}.

In this paper, we report that, with our improved orbit determination primarily using serendipitous archival observations as well as dedicated follow-up observations to extend the observed arc, we have identified 2019 UO$_{14}$ as the first known Trojan of Saturn.

\section{Observations}
\label{sec_obs}

\subsection{Follow-up Observation}
\label{ssec_fobs}

On UT 2024 April 4, we conducted a dedicated follow-up observation of 2019 UO$_{14}$ using the University of Hawai`i (UH) 2.2 m telescope on the summit of Maunakea, Hawai`i. Six individual exposures of 600 s tracked at the nonsidereal rate of the object were obtained with the STAcam imager. The images were $5 \times 5$ binned on chip to achieve the critical sampling of the realtime seeing on Maunakea (1\farcs4), rendering us an image scale of 0\farcs41 pixel$^{-1}$ and a dimension of $2112 \times 2112$ pixels. We subsequently calibrated the data with standard bias subtraction and flatfielding. The target 2019 UO$_{14}$ was clearly visible as a point source in all the individual exposures but the first two where it was completely clobbered by a field star. Therefore, the first two exposures are unusable. In the remaining images, the target exhibited an apparent motion in line with the ephemeris predictions used at that time. 

\subsection{Archival Observations}
\label{ssec_aobs}

In order to improve the orbit determination for 2019 UO$_{14}$, we exploited the Solar System Object Image Search (SSOIS) tool \citep{2012PASP..124..579G} at the Canadian Astronomy Data Centre (CADC) to search for archival observations of the object. We thereby managed to identify 2019 UO$_{14}$ in data taken by the Dark Energy Camera \citep[DECam;][]{2015AJ....150..150F} mounted at the V{\'i}ctor M. Blanco 4 m Telescope at Cerro Tololo Inter-American Observatory, Chile, the Hyper Suprime-Cam \citep[HSC;][]{2018PASJ...70S...1M} at the 8.2 m Subaru telescope, and the MegaCam camera \citep[][]{2003SPIE.4841...72B} at the 3.6 m Canada-France-Hawaii Telescope (CFHT), both on Manuakea. The serendipitous archival DECam data were taken through the $g$, $r$, and $i$ filters. Each of the 62 CCD chips covers a rectangular field of view (FOV) of $8\farcm9 \times 17\farcm7$ with an image scale of 0\farcs262 pixel$^{-1}$, while the overall covered sky region is approximately a hexagon of $\sim\!2\fdg2$ in diameter. We measured the FWHM of field stars that the seeing during these observations ranged from $\sim\!0\farcs9$ to 1\farcs8. The HSC, a mosaic camera attached at the prime focus of the Subaru telescope, consists of 104 main science CCD chips covering an overall circular FOV of $\sim\!1\fdg5$ in diameter in a pixel scale of $\sim\!0\farcs17$ pixel$^{-1}$ \citep{2018PASJ...70S...1M}. We identified 2019 UO$_{14}$ on one of the science CCD chips in two $g$-band HSC images both taken on UT 2019 November 7. Seeing significantly improved from $\sim\!1\farcs2$ in the first image to $0\farcs7$ in the second one. As for the CFHT/MegaCam data, they were all obtained through the $r$ filter from UT 2017 July 19-20. The camera is mosaicked by 40 CCD chips, each of which has a FOV of $6\farcm4 \times 14\farcm4$ under an image resolution of 0\farcs187 pixel$^{-1}$ in the binning $1 \times 1$ mode. Despite that all of the aforementioned observations were tracked sidereally, 2019 UO$_{14}$ does not appear to be visibly trailed therein thanks to its slow apparent motion. 

In addition, we searched for 2019\ UO$_{14}$ in the image archive of Pan-STARRS \citep[PS;][]{2016arXiv161205560C}, including {\it w}-band exposures.  Measurements were made of 209 images from mid-2015 to early 2022 which were survey exposures identified with an internal pre-discovery tool, and included remeasurements of all published observations in order to determine their astrometric uncertainties.  The two Pan-STARRS cameras have an image resolution of 0\farcs25 pixel$^{-1}$, and the FWHM of 2019\ UO$_{14}$ typically varied between 0\farcs8 and 1\farcs2 which matched the neighbouring stars.  Photometric measurements were made using the PS-calibrated zero-points in the observed filters ({\it w}, {\it i}, and {\it z} bands).

Other than the aforementioned serendipitous archival observations, with SSOIS we found dedicated $r$-band observations of 2019 UO$_{14}$ from the Gemini Multi-Object Spectrograph at the 8.1 m Gemini North telescope \citep[GMOS-N;][]{2004PASP..116..425H} atop Maunakea on UT 2021 February 13 and 16. The GMOS-N Hamamatsu CCD array is composed of three CCDs, providing a total field of FOV of $5\farcm5 \times 5\farcm5$. On-chip images of the observations were $2 \times 2$ binned, resulting in an image scale of $0\farcs16$ pixel$^{-1}$. We measured that seeing was $\sim\!1\farcs3$ on the first night, and on the second night, and that it varied slightly between $\sim\!0\farcs9$ and 1\farcs1. Despite that the telescope followed the apparent motion of the object, field stars did not appear to be noticeably trailed. This set of GMOS-N data was included for astrometry only. 

\begin{center}
\begin{figure*}[h!]
\epsscale{1.}
\plotone{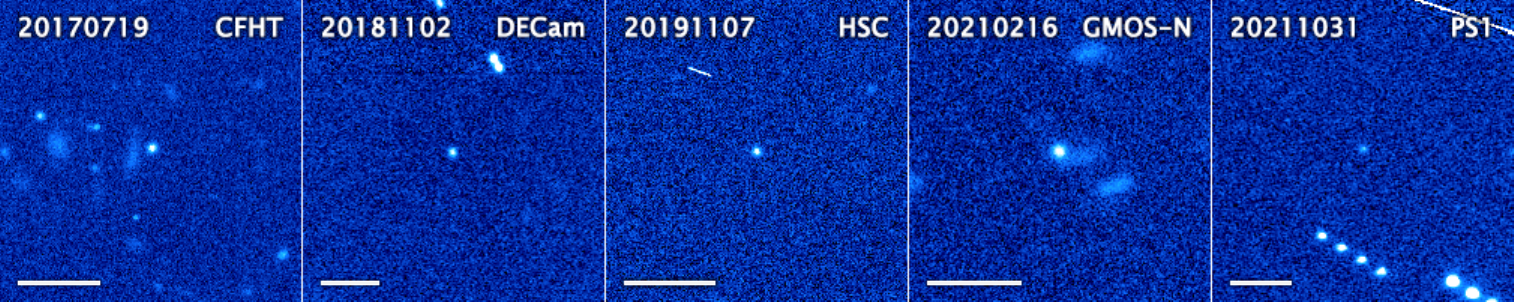}
\caption{Selected images of 2019 UO$_{14}$. All but the HSC image were coadded from individual exposures taken on the same nights. Each panel is annotated with the date of observation formatted as YYYYMMDD in the upper left corner and the used telescope/camera in the upper right corner. A horizontal white scale bar in the lower left corner represents an apparent length of 10\arcsec. The images are oriented such that the J2000 equatorial north is upwards, and east is to the left.}
\label{fig:obs}
\end{figure*}
\end{center}

Selected coadded images of 2019 UO$_{14}$ are shown in Figure \ref{fig:obs}, of which the target is at the center of each panel.

\section{Analysis}
\label{sec_ana}


\begin{deluxetable}{lcc}
\tablecaption{Best-fit Orbital Solution for 2019 UO$_{14}$
\label{tab:orb}}
\tablewidth{0pt}
\tablehead{
\multicolumn{2}{c}{Quantity}  & 
Value
}
\startdata
Semimajor axis (au) & $a$
       & 9.7956914(71) \\
Eccentricity & $e$
       & 0.23639022(48) \\ 
Inclination (\degr) & $i$
       & 32.8291884(28) \\ 
Longitude of perihelion (\degr) & $\varpi$
                 & 28.808263(32) \\
Argument of perihelion (\degr) & $\omega$
                 & 144.167981(32) \\ 
Longitude of ascending node (\degr) & ${\Omega}$
                 & 244.6402825(41) \\ 
Mean anomaly (\degr) & $M$
                  & 52.553994(69) \\
\hline
\multicolumn{2}{l}{Number of observations used (discarded)}
& 240 (0) \\
\multicolumn{2}{l}{Observed arc}
& 2015 Jun 26-2024 Apr 04 \\
\multicolumn{2}{l}{Residual rms (\arcsec)}
& 0.176 \\
\multicolumn{2}{l}{Normalized residual rms}
& 1.086
\enddata
\tablecomments{The osculating orbital elements are referred to the heliocentric J2000 ecliptic reference system at an epoch of TDT 2024 April 4.0 = JD 2460404.5. Here, we use the shorthand error notation to present the $1\sigma$ formal errors of orbital parameters.}
\end{deluxetable}

\subsection{Dynamics}
\label{ssec_dyn}

We carried out astrometric measurements for 2019 UO$_{14}$ using field stars and the Gaia DR2 catalogue \citep{2018A&A...616A...1G}. The slow apparent motion of the object allowed us to treat both the object and field stars as bidimensional circular Gaussians to be fitted, whereby we obtained the overall astrometric uncertainty propagated from errors in astrometric reduction and centroiding sources. Eight astrometric measurements from the Mount Lemmon Survey (G96) and two from the Lowell Discovery Telescope (G37) were identified and downloaded from the Minor Planet Center Explorer.\footnote{\url{https://data.minorplanetcenter.net/explorer/}} We then utilized the orbit determination package {\tt Find\_Orb}\footnote{The package was developed by B. Gray, freely available from \url{https://github.com/Bill-Gray/find_orb}.} to refine the orbital solution with these astrometric observations, debiased following the method detailed in \citet{2020Icar..33913596E} and weighted by the corresponding measurement errors. The solution took into account perturbations from the eight major planets, Pluto, the Moon, as well as the 16 most massive asteroids in the main belt, with their states determined from the planetary and lunar ephemeris DE440 \citep{2021AJ....161..105P}. In our preliminary orbital solution, we noticed that astrometric residuals of 23 astrometric measurements from PS exceeded their corresponding uncertainties beyond the $3\sigma$ level, even though all residuals were $\la\!0\farcs6$. As a result, we downweighted these observations according to their astrometric residuals and recalculated a best-fit solution for the orbit of 2019 UO$_{14}$, which yielded astrometric residuals comparable to the adopted uncertainties for all the astrometric observations we used. Table \ref{tab:orb} lists the best-fit Keplerian orbital elements along with their associated $1\sigma$ uncertainties, which were computed from the covariance matrix for the Keplerian orbital elements. Also included in Table \ref{tab:orb} is some fundamental information about the refined orbital solution.

With the refined orbital solution, we were ready for investigating the dynamical status of 2019 UO$_{14}$. We were most interested in knowing if the object would be a Saturn Trojan, in which case the key parameter to be examined is the mean longitude of the object relative to Saturn,
\begin{align}
\nonumber
\Delta l & \equiv l - l_{\rm S} \\
& = \left(\varpi + M \right) - \left(\varpi_{\rm S} + M_{\rm S}\right)
\label{eq_dl}.
\end{align}
\noindent Here, $l$, $\varpi$, and $M$ are the mean longitude, longitude of perihelion, and mean anomaly of 2019 UO$_{14}$, while those with the subscript `S' bear the same meanings yet for Saturn. If $\Delta l$ oscillates around $\pm60\degr$ with time, the object will be a Saturn Trojan. For $\Delta l$ oscillates around 0\degr, the object will be in a quasi-satellite of Saturn. If $\Delta l$ instead shows temporal oscillations around $\pm 180\degr$, it will be in a horseshoe orbit with Saturn. On the other hand, a circulating $\Delta l$ will simply indicate that the object is not a coorbital with Saturn.

\begin{figure}[ht!]
\epsscale{1.}
\begin{center}
\plotone{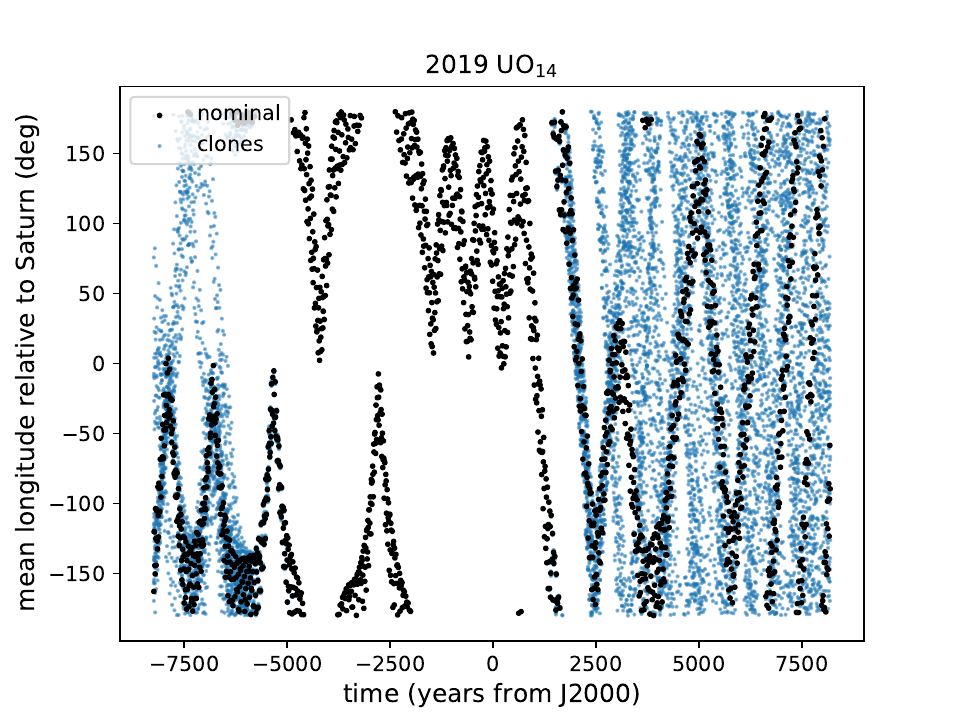}
\caption{The mean longitude of 2019 UO$_{14}$ (black) relative to Saturn along with 100 clones (blue) generated from the orbital covariance matrix.}
\label{fig:rellon}
\end{center}
\end{figure}

The past and future behavior of the mean longitude of 2014 UO$_{14}$ relative to Saturn is shown in Figure~\ref{fig:rellon}, along with 100 clones generated from our orbital covariance matrix. The integration was performed with a symplectic integrator \citep{1991AJ....102.1528W} able to handle close encounters with the planets using the Chambers formalism \citep{1999MNRAS.304..793C}, within a Solar System containing the eight planets, with the Earth and Moon combined into a single object at their mutual barycenter, and with initial conditions from the DE440 planetary ephemeris \citep{2021AJ....161..105P}. A time step of 3 days was used.

As long as the majority of the clones maintain the same type of behavior, we can be confident that the actual motion of 2019 UO$_{14}$ was as well; however once the clones begin to disperse, that indicates that behavior can only be understood statistically, as each clone is an equally likely instantiation of the object's true behavior within the orbital uncertainties.  From Figure~\ref{fig:rellon}, we see that the object has been in a Trojan state at the leading triangular Lagrange point $L_4$ for $\sim\!2$ kyr, which it entered from a horseshoe state. It is noteworthy that, prior to our work, two transient horseshoe coorbitals of Saturn, (15504) 1999 RG$_{33}$ and 2013 VZ$_{70}$, have been identified \citep{2006Icar..184...29G,2021PSJ.....2..212A,2022A&A...657A..59D}. At times earlier than $\sim\!6$ kyr ago, the clones disperse and the precise motion can no longer be determined. Looking to the future, the Trojan state will be maintained for about 1 kyr, before transitioning to a horseshoe state for roughly a thousand years more. The measured Lyapunov time of 2019 UO$_{14}$ is $\sim\!0.5$~kyr, consistent with the rates of clone dispersal.

\begin{figure}[ht!]
\epsscale{1.}
\begin{center}
\plotone{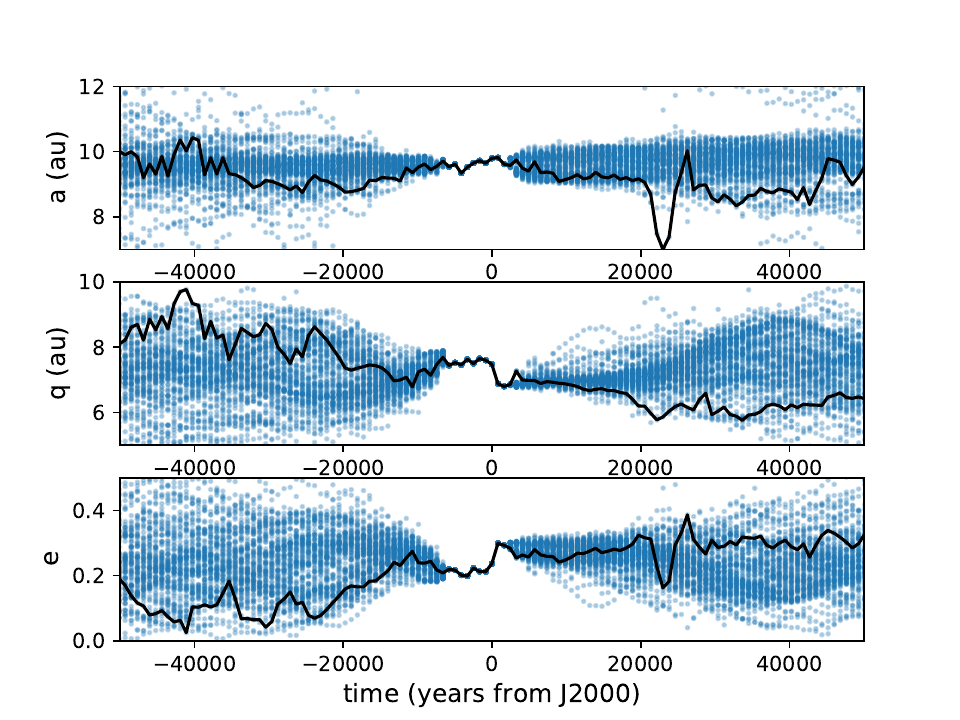}
\caption{The time evolution of the semimajor axis $a$, perihelion distance $q$, and eccentricity $e$ of the nominal orbit of 2019 UO$_{14}$ (black) and 100 clones (blue).}
\label{fig:orbevo}
\end{center}
\end{figure}

Looking further into the past, the majority of the clones remain in one or another co-orbital state back for 15 kyr at which point the resonance is broken. The orbit from which 2019 UO$_{14}$ entered into the coorbital state is consistent with that of the background Centaurs, and therefore the object was likely captured from this population and is currently being trapped as a transient Saturn Trojan. Our finding that 2019 UO$_{14}$ is a short-lived Saturn Trojan is in line with previous dynamical studies, as the object lies in the strongly unstable region for Saturn Trojans, where the orbit will be quickly destabilized by Jupiter \citep[e.g.,][]{2002Icar..160..271N}.

In the current state, our numerical integration simulation exhibits that 2019 UO$_{14}$ librates around $L_4$ in a period of $\sim\!0.7$ kyr, which is in excellent agreement with the value yielded by the formula \citep{1999ssd..book.....M}:
\begin{equation}
T_{\rm L} = \frac{4\pi}{3}\sqrt{\frac{a_{\rm S}^3}{3 \mu_{\rm S}}}
\label{eq_plib}.
\end{equation}
\noindent Here, $a_{\rm S}$ is the semimajor axis of Saturn's heliocentric orbit, and $\mu_{\rm S}$ is the mass parameter of Saturn. Substituting $a_{\rm S} = 9.6$ au and $\mu_{\rm S} = 8.5 \times 10^{-8}$ au$^{3}$ d$^{-2}$, we recover $T_{\rm L} \approx 0.7$ kyr.

At present 2019 UO$_{14}$ is nevertheless a Saturn Trojan, despite that it is a short-lived one. Therefore, all of the four giant planets in our solar system have Trojan populations, while Mercury and Venus remain the only two major planets without any known associated Trojans.



In addition to investigating the Trojan status of 2019 UO$_{14}$, we examined the temporal evolution of its orbit in terms of the semimajor axis $a$, perihelion distance $q$, and eccentricity $e$ over an extended timeframe, from 50 kyr ago to 50 kyr in the future (Figure \ref{fig:orbevo}). The orbital chaoticity is evident, with the timescale of the clone dispersal in line with the measured Lyapunov time. While it is inappropriate to use the past orbital evolution of the nominal orbit and its clones too much beyond the Lyapunov time to infer the source region of 2019 UO$_{14}$, it is valid to statistically assess the future orbital evolution. As seen in Figure \ref{fig:orbevo}, except for the nominal orbit and a few clones, the semimajor axis remains largely stable without significant jumps $\ga\!2$ au in the next 50 kyr. However, the dispersion in the perihelion distance $q$ of the object appears to be more pronounced, as a result of the varying eccentricity $e$. It seems more probable that the perihelion distance will increase over the next $\sim\!40$ kyr. Yet, the possibility of further decrease in the perihelion distance is not negligible either, which is the case for the nominal orbit. While 2019 UO$_{14}$ is more likely to possess Centaur-like orbits in the next 50 kyr, we also notice that a small fraction ($\la\!10\%$) of the clones will evolve into orbits typical of Jupiter-family comets.

\subsection{Physical Properties}
\label{ssec_phys}

\begin{figure}
\begin{center}
\gridline{\fig{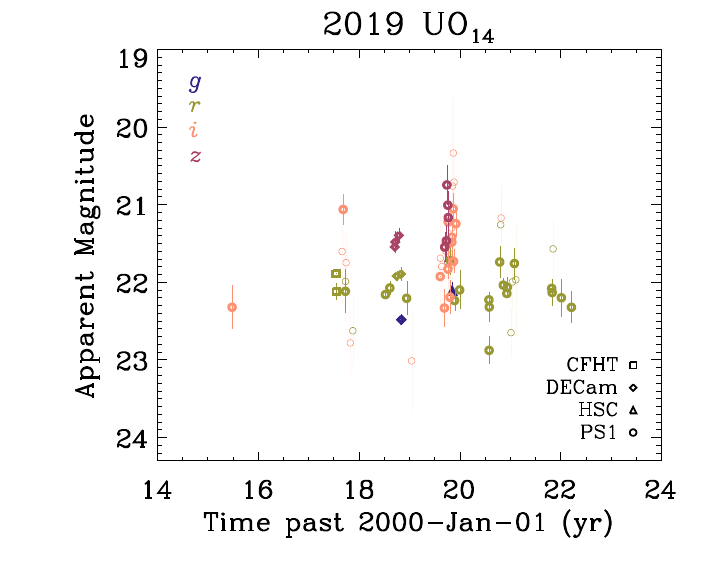}{0.5\textwidth}{(a)}
\fig{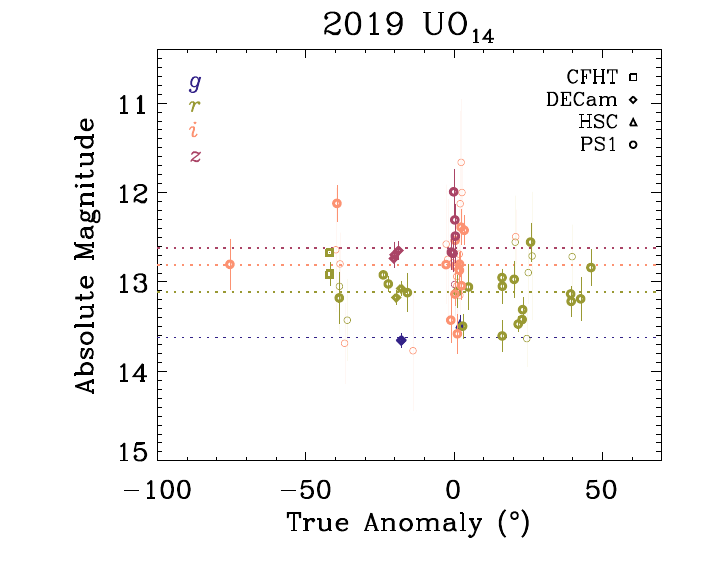}{0.5\textwidth}{(b)}}
\caption{
Apparent magnitude of 2019 UO$_{14}$ as a function of time (a) and its absolute magnitude versus true anomaly (b). The horizontal dotted lines in panel (b) represent weighted means of absolute magnitude in the corresponding bands, which are distinguished by colors. As the legends indicate, data points from different telescopes are plotted as different symbols. To improve visual clarity, measurements with uncertainties $\le 0.3$ mag are plotted in bolder font.
\label{fig:mag}
} 
\end{center}
\end{figure}

\begin{figure}
\epsscale{1.}
\begin{center}
\plotone{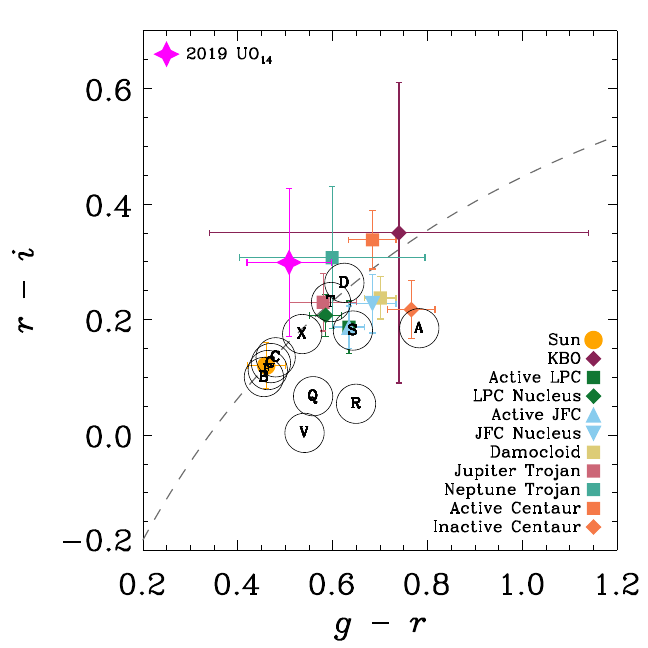}
\caption{
The color of 2019 UO$_{14}$ (plotted as star in magenta) in comparison to those of miscellaneous populations of small solar system bodies \citep[and citations therein]{2003Icar..163..363D,2007MNRAS.377.1393S,2012Icar..218..571S,2015AJ....150..201J,2023PSJ.....4..135M} and the Sun \citep{2018ApJS..236...47W} in the SDSS $g-r$ and $r-i$ regimes. Open circles with letters labelled mark typical colors of main-belt asteroid taxonomic classes. We followed \citet{2006A&A...460..339J} and applied transformations for reported color indices not in the SDSS system. The grey dashed curve represents the locus of bodies with linear reflectivity spectra in the investigated space. We notice that the measured color of 2019 UO$_{14}$ is in line with those of Jupiter and Neptune Trojans, and that it is less similar to those of Centaurs, yet not beyond the $3\sigma$ level.
\label{fig:clr}
} 
\end{center} 
\end{figure}

We conducted aperture photometry for 2019 UO$_{14}$ in the archival observations from CFHT, DECam, HSC, and PS1 using multiple circular apertures ranging from 2\arcsec~to 5\arcsec~in radius, with a step size of 0\farcs5. Images from the former two facilities were first photometrically calibrated with the ATLAS Refcat2 catalogue \citep{2018ApJ...867..105T} and transformed into the SDSS photometric system from the PS1 system according to \citet{2012ApJ...750...99T}. The obtained apparent magnitude of the object in the SDSS system from the 2\arcsec~radius aperture versus time is shown in Figure \ref{fig:mag}a. We also examined results from larger apertures, finding that while they produced visually similar plots, they also exhibited progressively greater errors and scatter as the aperture size increased. As a result, we decided to focus on measurements obtained from the 2\arcsec~radius aperture only. 

We suspected that the varying viewing geometry might have played a role in causing the scatter in Figure \ref{fig:mag}a, in addition to the low signal-to-noise ratios (S/N) of 2019 UO$_{14}$ in the serendipitous archival observations. To eliminate this factor, we computed the absolute magnitude through
\begin{equation}
H_{\lambda} = m_{\lambda} - 5 \log r_{\rm H} \Delta - \beta_{\alpha} \alpha
\label{eq_mabs},
\end{equation}
\noindent where $m_{\lambda}$ and $H_{\lambda}$ are, respectively, the apparent and absolute magnitudes in a given bandpass $\lambda$, $r_{\rm H}$ and $\Delta$ are, respectively, heliocentric and target-observer distances, $\alpha$ is phase angle, and $\beta_{\alpha}$ is the linear phase coefficient. We assumed $\beta_{\alpha} = 0.06 \pm 0.01$ mag deg$^{-1}$ as per values reported for Centaurs and Jupiter Trojans \citep[e.g.,][]{2007AJ....133...26R,2010Icar..207..699S,2023PSJ.....4...75D}, which is also consistent with the value adopted by \citet{2020AJ....159..209L}. Figure \ref{fig:mag}b presents the obtained absolute magnitude plotted against true anomaly of 2019 UO$_{14}$. Unfortunately, significant scatter is still observed in the $r$-, $i$-, and $z$-band data points, indicating that the primary cause of the scatter is the low S/N of the object in the observations. As for the $g$-band measurements, there were only two nights of data. 2019 UO$_{14}$ has never been fortuitously observed in more than a single filter by any of the aforementioned facilities on the same nights. In light of these issues, we computed weighted means for absolute magnitudes in the bandpasses from the corresponding repeated measurements (over-plotted as dotted lines in Figure \ref{fig:mag}). The results are $H_{g} = 13.62 \pm 0.05$ (0.08), $H_{r} = 13.11 \pm 0.07$ (0.27), $H_{i} = 12.81 \pm 0.10$ (0.39), and $H_{z} = 12.63 \pm 0.09$ (0.21), where the unbracketed and bracketed errors are, respectively, errors and standard deviations of the weighted means. Assuming a nominal $r$-band geometric albedo of $p_r = 0.05$, the equivalent radius of 2019 UO$_{14}$ is $R_{\rm n} = 10^{0.2 \left(m_{\odot,r} - H_r\right)} r_{\oplus} / \sqrt{p_r} = 6.6 \pm 0.2$ km, where $m_{\odot,r} = -26.93 \pm 0.03$ is the $r$-band apparent magnitude of the Sun at $r_{\oplus} = 1$ au \citep{2018ApJS..236...47W}.

We were then able to obtain the color of 2019 UO$_{14}$ to be $g - r = +0.51 \pm 0.09$, $r - i = +0.30 \pm 0.13$, $i - z = +0.19 \pm 0.14$, in which the uncertainties were propagated from errors on the weighted means. Figure \ref{fig:clr} presents a comparison in the $g-r$ vs. $r-i$ space between the color of 2019 UO$_{14}$ and those of various populations of small solar system objects, along with the solar color. We thereby immediately noticed that 2019 UO$_{14}$ has a color in the $g-r$ and $r-i$ regimes closest to the counterparts of Jupiter and Neptune Trojans reported in \citet{2007MNRAS.377.1393S},  \citet{2012Icar..218..571S}, and \citet{2023PSJ.....4..135M}. Given the uncertainty, its color can also potentially resemble blue Kuiper-belt objects (KBOs). When compared to Centaurs, the color appears to be less similar. However, the discrepancy does not exceed the $3\sigma$ level, and therefore, it is not statistically confident to conclude that the color of the Saturn Trojan is different from those of Centaurs.


\begin{figure*}
\begin{center}
\gridline{\fig{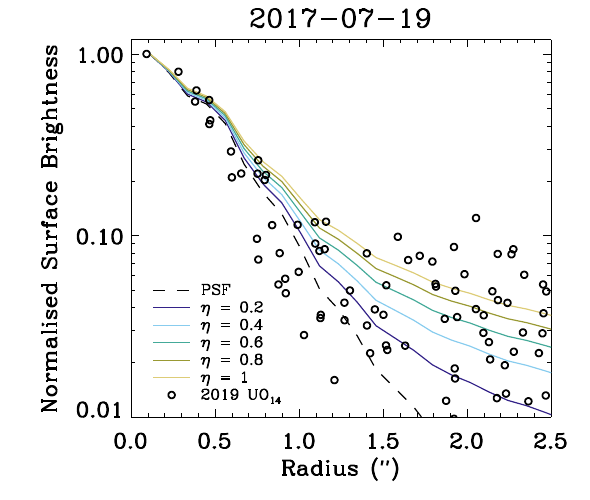}{0.5\textwidth}{(a)}
\fig{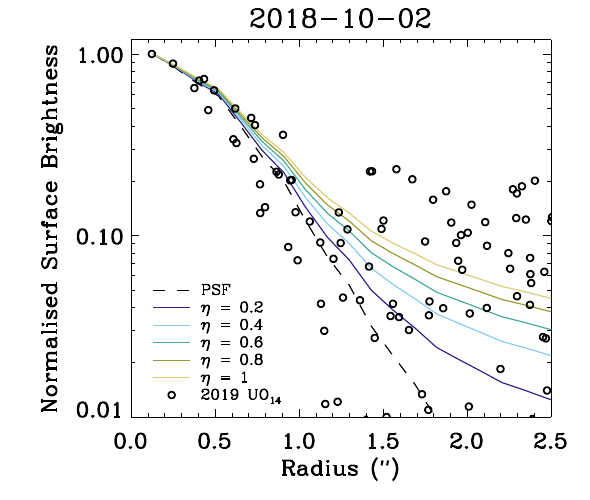}{0.5\textwidth}{(b)}
}
\gridline{
\fig{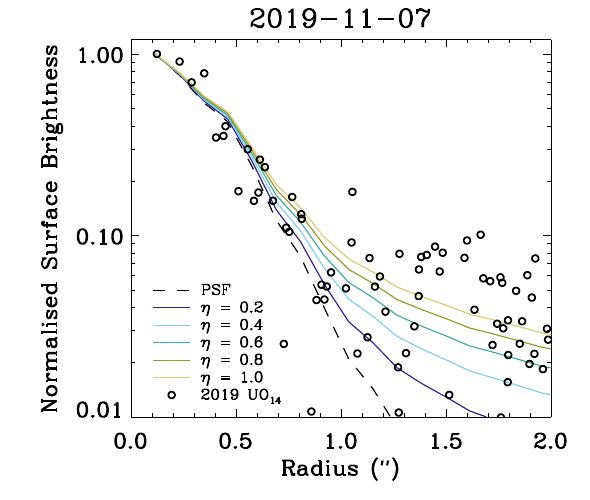}{0.5\textwidth}{(c)}}
\caption{Comparison of the normalized radial surface brightness profile of 2019 UO$_{14}$ (open dots) with those of the PSF (black dashed curve) and models of varying activity levels (solid curves in various colors, as indicated in the legends) on (a) 2017 July 19, (b) 2018 October 2, and (c) 2019 November 7. Despite that these examples have the best S/N if compared to other observations, the signal from 2019 UO$_{14}$ becomes dominated by the sky background beyond $\sim\!1\arcsec$ from the optocenter. Taking the scatter into consideration, no obvious differences can be discerned in the radial brightness profiles between the object and the PSF.}
\label{fig:rprof}
\end{center} 
\end{figure*}

Finally, we attempted to examine whether the Saturn Trojan was active in the $r$-band archival data from CFHT and DECam, where the object achieved better S/N. If the object was active, the flux contribution from its dust component would surpass that from its gas component, due to the dominance of the scattering cross-section from dust grains. After conducting a visual inspection and finding no apparent cometary features, we proceeded to calculate and compared its FWHM to the those of the point-spread functions (PSFs) extracted from the data using field stars with {\tt StarFinder} \citep{2000A&AS..147..335D,2000SPIE.4007..879D}. Our result was that 2019 UO$_{14}$ maintained a FWHM highly consistent with the PSF in the examined images. Therefore, we can conclude that the available observations provide no compelling evidence of the object being active.  A similar analysis was performed on the PS1 images, which also did not provide evidence of any activity.

To establish an upper limit for the activity of 2019 UO$_{14}$, we followed the method by \citet{1992Icar...97..276L} and assumed a synthesized steady-state coma with varying fractional contributions. Thus, the surface brightness profile of such a model can be expressed as
\begin{equation}
\mathcal{S}\left({\bf r}\right) = \left[k_{\rm n} \delta\left({\bf r}\right) + \frac{k_{\rm c}}{r}\right] \ast \mathcal{P}\left({\bf r}\right)
\label{eq_sb},
\end{equation}
\noindent where $k_{\rm n}$ and $k_{\rm c}$ are scaling coefficients for the nucleus and the coma, respectively, $\delta$ is the Dirac delta function, symbol $\ast$ denotes the operation of convolution, and $\mathcal{P}$ is the PSF expressed in polar coordinates ${\bf r} = \left(r,\theta\right)$, with its total intensity normalized to unity. Note also that the above apparent radius $r$ from the nucleus should be expressed in radians. Therefore, the parameter $\eta$, which is the flux ratio between the coma and the nucleus within the circular aperture of radius $r_0$, is then
\begin{align}
\nonumber
\eta & = \frac{\iint_{S} k_{\rm c} r^{-1} {\rm d}S}{\iint_{S} k_{\rm n} \delta\left({\bf r}\right) {\rm d}S} \\
& = \frac{2\pi k_{\rm c} r_{0}}{k_{\rm n}}
\label{eq_eta}.
\end{align}
\noindent Here, $S$ represents the aperture area. In Figure \ref{fig:rprof}, we compare the normalized radial brightness profiles of 2019 UO$_{14}$, the PSF, and models with different fractional contributions of the dust coma (parameterized by $\eta$) using images from CFHT taken on 2017 July 19, DECam taken on 2018 October 2, and HSC taken on 2019 November 7. In these instances where the object achieved S/N higher than other observations, the radial profile of the object is indistinguishable from the PSF, given the scatter of the object's signal. Beyond $\sim\!1\arcsec$ from the optocenter, the scatter worsens notably as the sky background begins to dominate. By comparing the radial profiles of 2019 UO$_{14}$ and the models within 2\arcsec~from the optocenter in Figure \ref{fig:rprof}, we conservatively set an upper limit of $\eta \la 1$ for the fractional contribution of the dust coma. This allows us to further estimate the optical depth of the dust coma within the aperture to be $\tau = \eta R_{\rm n}^{2} / \left(r_{0} \Delta \right)^2 \la 4 \times 10^{-7}$. For comparison, the dust trail of Centaur 29P/Schwassmann-Wachmann 1 was reported to have an optical thickness of $\tau \sim 10^{-8}$ by \citet{2004ApJS..154..463S}, while the gossamer rings of Jupiter have $\tau \sim 10^{-7}$ \citep{2018prs..book..125D}.

Assuming dust grains drift radially from the nucleus at a constant speed in steady state, the mass-loss rate of dust can be calculated as the total dust mass within the circular aperture of radius $r_0$, divided by the time it takes for dust grains to travel from the nucleus to the edge of the aperture. We thus derived
\begin{equation}
\dot{M}_{\rm d} = \frac{4 \pi \eta \rho_{\rm d} \bar{\mathfrak{a}}_{\rm d} \bar{v}_{\rm d} R_{\rm n}^2}{3 r_{0} \Delta}
\label{eq_mdot},
\end{equation}
\noindent in which $\rho_{\rm d}$, $\bar{\mathfrak{a}}_{\rm d}$, and $\bar{v}_{\rm d}$ are the bulk density, mean radius, and mean speed of dust grains, respectively, $r_0$ is the aperture radius expressed in radians, and $S$ denotes the projected area of the circular aperture centered on the nucleus. Substituting the obtained values into Equation (\ref{eq_mdot}) yields
\begin{equation}
\dot{M}_{\rm d} \la \left(17~\text{kg s$^{-1}$}\right) \left(\frac{\rho_{\rm d}}{1~\text{g cm$^{-3}$}}\right) \left(\frac{\bar{\mathfrak{a}}_{\rm d}}{1~\text{mm}} \right) \left(\frac{\bar{v}_{\rm d}}{1~\text{m s$^{-1}$}} \right)
\label{eq_mdot_sub}.
\end{equation}
\noindent Assuming the physical properties of dust grains of 2019 UO$_{14}$ similar to those of 29P reported in \citet{1992Natur.359...42F} and \citet{2022A&A...664A..95B}, we obtained a dust mass-loss rate of $\dot{M}_{\rm d} \la 1$ kg s$^{-1}$ from Equation (\ref{eq_mdot_sub}), which is at least an order of magnitude smaller than that of 29P during its quiescent phase \citep{2022A&A...664A..95B}. Our constraint on the activity of 2019 UO$_{14}$ is comparable to those of the Centaurs studied by \citet[and citations therein]{2020AJ....159..209L} at similar heliocentric distances. 

\begin{figure*}[h!]
\begin{center}
\gridline{\fig{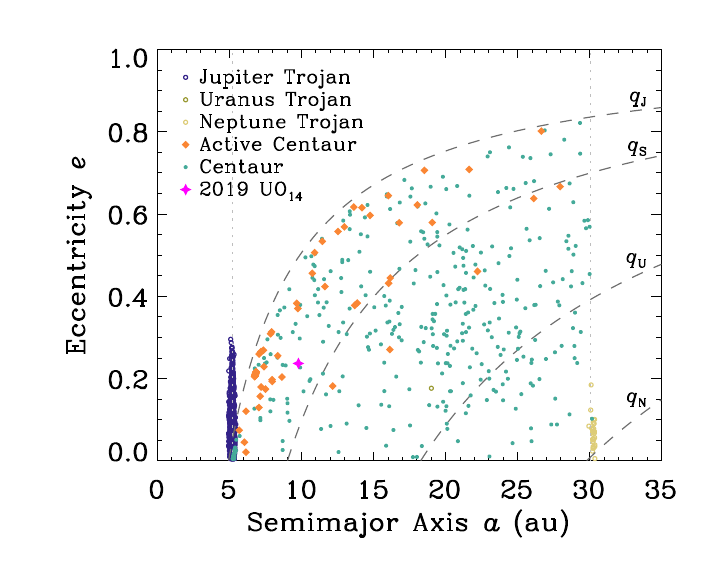}{0.5\textwidth}{(a)}
\fig{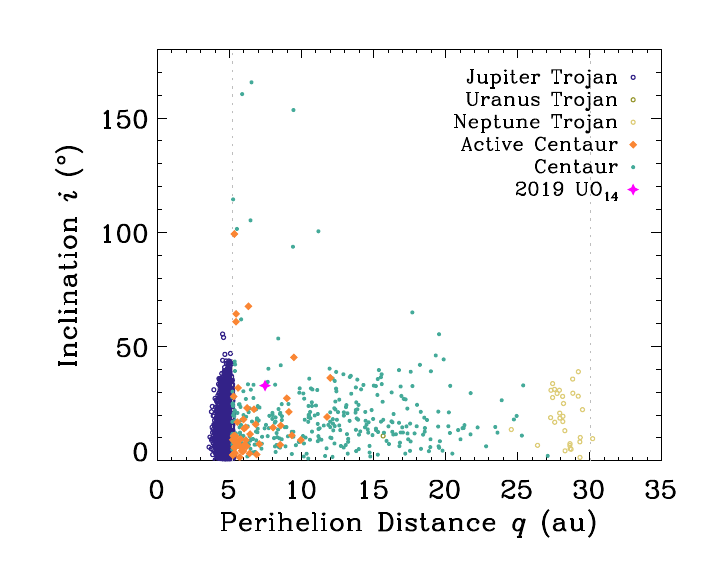}{0.5\textwidth}{(b)}}
\caption{
Comparison of 2019 UO$_{14}$ (plotted as a magenta star) to other Trojans and Centaurs, with active Centaurs highlighted (see the legends), in the space of (a) semimajor axis ($a$) versus eccentricity ($e$) and (b) perihelion distance ($q$) versus inclination ($i$). Other Trojans were queried from the MPC, and Centaurs were sourced from the JPL Small-Body Database Query with constraints on semimajor axes $a_{\rm J} \le a \le a_{\rm N}$, perihelion distances $a_{\rm J} \le q \le a_{\rm N}$, and exclusions of Trojans ($a_{\rm J}$ and $a_{\rm N}$ are the semimajor axes of Jupiter and Neptune, respectively). The list of active Centaurs was compiled using \citet{2009AJ....137.4296J} and \citet{2012AJ....144...97G}, along with objects with cometary designations from the JPL Small-Body Database Query. The four labelled dashed curves in the left panel represent loci of orbits with perihelion distances equal to those of the four giant planets. The two vertical grey dotted lines in both panels mark the boundaries of the Centaur region at the semimajor axes of Jupiter and Neptune.
\label{fig:orb_comp}
}
\end{center}
\end{figure*}

Despite the current observations showing no compelling evidence of the activity in 2019 UO$_{14}$, we strongly advocate for future observations to search for its potential cometary activity. We argue that this effort is worthwhile and not on a wild goose chase. To support this, we estimate the average surface temperature of the object over a full orbital period $P$ (from arbitrary initial time $t_0$ to end time $t_0 + P$) using the energy equilibrium equation that balances insolation and thermal reradiation, deriving
\begin{align}
\nonumber
\bar{T}_{\rm s} & = \frac{1}{2}\left[\frac{\left(1-A_{\rm B}\right)L_{\odot}}{\pi \epsilon \sigma_{\rm B} P} \int\limits_{t_{0}}^{t_{0} + P} \frac{{\rm d}t}{r_{\rm H}^{2}}\right]^{1/4} \\
& = \frac{1}{2} \left[\frac{\left(1-A_{\rm B}\right) L_{\odot}}{\pi \epsilon \sigma_{\rm B} a^2 \sqrt{1 - e^2}} \right]^{1/4}
\label{eq_Ts}.
\end{align}
\noindent Here, $A_{\rm B} \sim 10^{-2}$ is the Bond albedo, $\epsilon \sim 1$ is the emissivity, $L_{\odot} = 3.8 \times 10^{26}$ W is the solar luminosity, and $\sigma_{\rm B} = 5.67 \times 10^{-8}$ W m$^{-2}$ K$^{-4}$ is the Stefan-Boltzmann constant. Substituting the values of the semimajor axis and eccentricity in Table \ref{tab:orb}, we obtain $\bar{T}_{\rm s} \approx 90$ K. The timescale for crystallization of amorphous ice, an exorthermic transition, is sensitive to the ambient temperature $T$ as:
\begin{equation}
\tau_{\rm cr} = \tau_{{\rm cr},0} \exp\left(-\frac{E_{\rm A}}{k_{\rm B} T}\right)
\label{eq_tau_cr},
\end{equation}
\noindent where $\tau_{{\rm cr},0} = 3.0 \times 10^{-21}$ yr is a scaling coefficient, $k_{\rm B}$ is the Boltzmann constant, and $E_{\rm A}$ is the activation energy with $-E_{\rm A} / k_{\rm B} = 5370$ K \citep{1989ESASP.302...65S}. Equating the ambient temperature to $\bar{T}_{\rm s}$ from Equation (\ref{eq_Ts}), we find $\tau_{\rm cr} \approx 8 \times 10^4$ yr, two orders of magnitude longer than the measured Lyapunov time of 2019 UO$_{14}$. Given that the object was likely captured from the Centaur region, this suggests that it may still retain amorphous ice. Additionally, since the perihelion distance of 2019 UO$_{14}$ is intermediate between the those of active and inactive Centaurs, and cometary activity has been previously observed in some Centaurs at similar distances from the Sun \citep[][and see also Figure \ref{fig:orb_comp}]{2009AJ....137.4296J,2012AJ....144...97G,2020AJ....159..209L}, we postulate that 2019 UO$_{14}$ has a promising potential to be an active Trojan. If future observations confirm its cometary activity, this would mark the object as the first active Trojan in our solar system. Conversely, if the object remains inactive, it may indicate a depletion of its amorphous ice, hinting at a history of the present-time Saturn Trojan as a Jupiter-family comet. Nonetheless, the residence of 2019 UO$_{14}$ in the crystallization zone significantly simplifies the probing of its past evolution compared to small bodies from other regions. As such, we anticipate that future studies involving a larger population of Saturn Trojans, in particular the primordial ones, will facilitate the imposition of much more stringent constraints on the formation and evolution of the solar system.

\section{Summary}
\label{sec_sum}

In this paper, we performed a dynamical and photometric analyses of 2019 UO$_{14}$ primarily using serendipitous archival and followup observations. The key results are listed as follows:
\begin{enumerate}
    \item We identified the object as the first Trojan of Saturn, librating around $L_4$ of the Sun-Saturn system in a period of $\sim\!0.7$ kyr. Therefore, all of the four giant planets in the solar system have their Trojan populations.

    \item However, our $N$-body integration revealed that the object is only a transient Saturn Trojan, as it was likely captured from the Centaur population $\sim\!2$ kyr ago from a horseshoe coorbital and will restore its horseshoe state roughly a millennium later.

    \item Assuming a linear phase coefficient of $0.06 \pm 0.01$ mag/deg as appropriate for Centaurs and Jupiter Trojans, we measured the $r$-band absolute magnitude of the object to be $H_r = 13.11 \pm 0.07$. Adopting an $r$-band geometric albedo of $p_r = 0.05$, the intrinsic brightness requires an effective radius of the nucleus of $6.6 \pm 0.02$ km.

    \item We obtained the color of the object to be $g - r = +0.54 \pm 0.07$, $r - i = +0.30 \pm 0.13$, and $i - z = +0.19 \pm 0.14$. The object has a color in the $g-r$ vs. $r-i$ space closely resembling Jupiter and Neptune Trojans, potentially similar to blue KBOs, and not statistically different from Centaurs, after the uncertainty taken into consideration.

    \item The object exhibited no compelling evidence of being active in any of the observations, as its FWHM remained consistent with those of the PSFs. The most stringent constraint on the activity of the object by means of comparing radial brightness profiles is that the optical depth of dust within our photometric aperture was $\la\!10^{-7}$. Assuming the physical properties of dust grains are similar to those of Centaur 29P/Schwassmann-Wachmann 1, we found that the upper limit to the mass-loss rate of dust was $\la\!1$ kg s$^{-1}$. 

    \item Despite the current lack of detected activity, we highlight the possibility that the object could be an active Trojan. We postulate that future observations may reveal signs of its activity. If confirmed, this would mark the object as the first active Trojan in our solar system.
\end{enumerate}

\begin{acknowledgements}

We thank the anonymous reviewers for their comments on the manuscript, Bill Gray for making his orbit determination package {\tt Find\_Orb} publicly available, and people from G37 and G96 who submitted their astrometry of 2019 UO$_{14}$ to the Minor Planet Center. The work was supported by the Science and Technology Development Fund, Macau SAR, through grant Nos. 0051/2021/A1 and 0016/2022/A1 to M.T.H.

\end{acknowledgements}
\vspace{5mm}

\software{{\tt Find\_Orb}, {\tt StarFinder} \citep{2000A&AS..147..335D,2000SPIE.4007..879D}}


\bibliography{ST_bib}

\end{document}